\documentclass[amssymb,prb,twocolumn,showpacs,superscriptaddress]{revtex4}
\usepackage{epsfig}
\usepackage{dcolumn}
\usepackage{amsmath,amssymb}

\begin{document}

\title{Magnetic-field instability of Majorana modes in multiband quantum wires} 

\author{Jong Soo Lim}
\affiliation{Institut de F\'{\i}sica Interdisciplin\`aria i de Sistemes Complexos
IFISC (CSIC-UIB), E-07122 Palma de Mallorca, Spain}
\author{Lloren\c{c} Serra}
\affiliation{Institut de F\'{\i}sica Interdisciplin\`aria i de Sistemes Complexos
IFISC (CSIC-UIB), E-07122 Palma de Mallorca, Spain}
\affiliation{Departament de F\'{\i}sica,
Universitat de les Illes Balears, E-07122 Palma de Mallorca, Spain}
\author{Rosa L\'opez}
\affiliation{Institut de F\'{\i}sica Interdisciplin\`aria i de Sistemes Complexos
IFISC (CSIC-UIB), E-07122 Palma de Mallorca, Spain}
\affiliation{Departament de F\'{\i}sica,
Universitat de les Illes Balears, E-07122 Palma de Mallorca, Spain}
\author{Ram\'on Aguado}
\affiliation{Teor\'{\i}a y Simulaci\'on de Materiales,
Instituto de Ciencia de Materiales de Madrid, ICMM-CSIC Cantoblanco,
E-28049 Madrid, Spain}

\date{May 2, 2012}

\begin{abstract}
We investigate the occurrence of Majorana modes in semiconductor 
quantum wires in close proximity with a superconductor and when both, 
Rashba interaction and magnetic field, are present. 
We consider long, but finite, multiband wires (namely, planar wires
with dimensions $L_x\gg L_y$). Our results demonstrate that interband mixing
coming from Rashba spin orbit term hybridizes Majorana pairs originating from different
transverse modes while simultaneously closing the effective gap. Consequently, multiple  Majorana modes
do not coexist in general. On the contrary, Majorana physics is robust
provided that only one single transverse mode contributes with a
Majorana pair. Finally, we analyse the robustness of  Majorana physics with respect to magnetic orbital effects.
\end{abstract}

\maketitle

\emph{Introduction.} Matter and its charge conjugate counterpart obey the relativistic  Dirac equation with positive and negative energies, respectively. \cite{Dirac}  Ettore Majorana proved in 1937 that these solutions 
exist for which particle and antiparticle are the same entity, a \textit{Majorana fermion}.\cite{Majorana}  Majorana fermions 
were first proposed in the context of particle physics to describe neutrinos \cite{PhysRevLett.9.36} 
and, more recently, the Majorana search has been revived \cite{Majorana-short1} in the condensed matter,  
\cite{PhysRevB.44.9667,springerlink:10.1134/1.568223,PhysRevB.61.10267,
Moore1991362,PhysRevB.73.220502,PhysRevLett.100.096407,PhysRevB79.161408,PhysRevB.81.125318,PhysRevLett.105.077001,
PhysRevLett.105.177002,PhysRevB.82.085314,PhysRevB.84.081304} and atomic physics  
\cite{PhysRevLett.91.090402,PhysRevLett.95.070404,PhysRevLett.98.010506,PhysRevLett.106.220402} communities. 
Aside from a fundamental interest in finding Majorana fermions, 
part of the excitement comes from the non-Abelian braiding statistics
appearing when these particles are localized near a vortex or a domain wall, 
which could be useful for topological quantum computation. \cite{Majorana-short2}

The most recent and promising proposal for engineering  Majorana quasiparticles is based on semiconductor nanowires (NWs) put in close proximity with a superconductor when both spin orbit (SO) interaction and magnetic field are present. The 
main advantages of these systems are that they exhibit giant Zeeman splittings due to the huge g-factor and that conventional s-wave superconductivity can be proximity induced. \cite{Nadj-Perde1,Lund1,Lund2}  Majorana modes are quasiparticles excitations that emerge when the Kramers degeneracy of the electron-hole pairs is lifted  because of the presence of both Rashba SO and magnetic field.  Nontrivial phases that correspond to the occurrence of Majorana physics have been predicted to survive even in multimode NWs  provied that the number of filled subbands is odd. \cite{PhysRevLett.106.127001,Lutchyn-Fisher,PhysRevB.83.184520,PhysRevB.84.144522,Fazio}  In this manner, 
 trivial and non-trivial topological phases can be alternated by tuning either the chemical potential or the magnetic field. 
 Nevertheless, this scenario of alternating robust (nontrivial) and fragile (trivial) phases is based on topological arguments 
 strictly valid for \emph{infinite} long strips that are characterized by the $Z_2$ invariant, such as the Majorana number. \cite{Kitaev01}  In order to unambiguously predict the occurrence of Majorana 
 modes in realistic systems, one needs to investigate \emph{finite} multimode nanowires.
Such studies have been previously attempted in two-band model and tight-binding Hamiltonians. 
\cite{PhysRevB.83.184520,PhysRevB.84.144522,PhysRevLett.106.127001,PhysRevB.85.060507} Here, we perform \emph{exact} 
numerical diagonalization of the Hamiltonian. Our main finding indicates that multiple Majorana modes in finite multiband nanowires cannot coexist due to the strong hybridization caused by the  Rashba spin orbit interaction. The importance 
of the Rashba coupling term in planar wires has been emphasized by some of us. \cite{ric1,ric2,ric3}  
We observe the formation of a single Majorana pair at low magnetic fields only. 
Furthermore,  we investigate the magnetic orbital effects in these systems and 
we show that the single Majorana pair regime survives \emph{solely}  if the magnetic 
field is in the plane of the wire.
\begin{figure}[t]
\centerline{
\epsfig{file=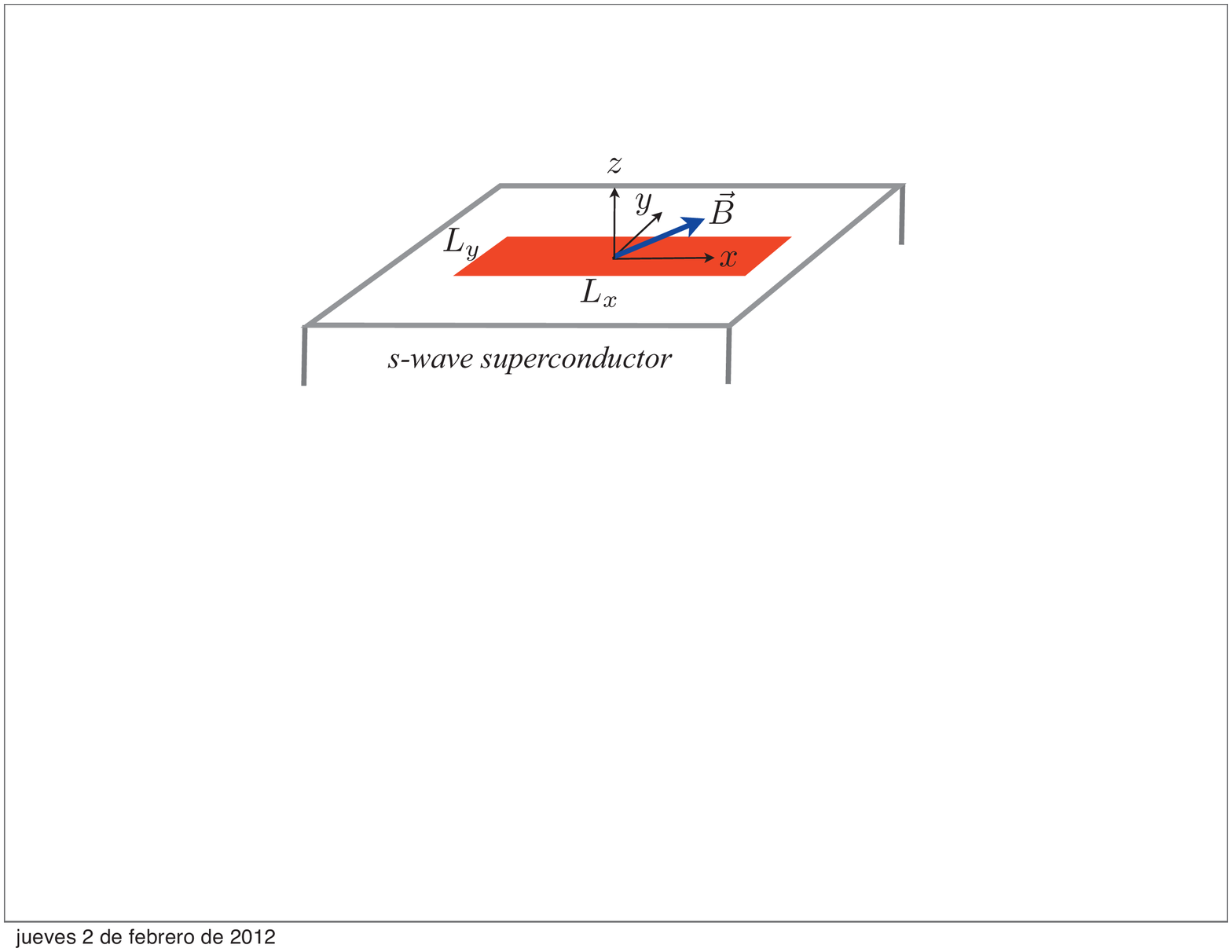,angle=0,width=0.35\textwidth,clip}
}
\caption{(Color online) Sketch of a finite nanowire of dimensions
$(L_x,L_y)$ in close proximity with an s-wave superconductor and in presence of a tilted magnetic field.}
\label{fig1}
\end{figure}

\begin{figure}[t]
\centerline{
\epsfig{file=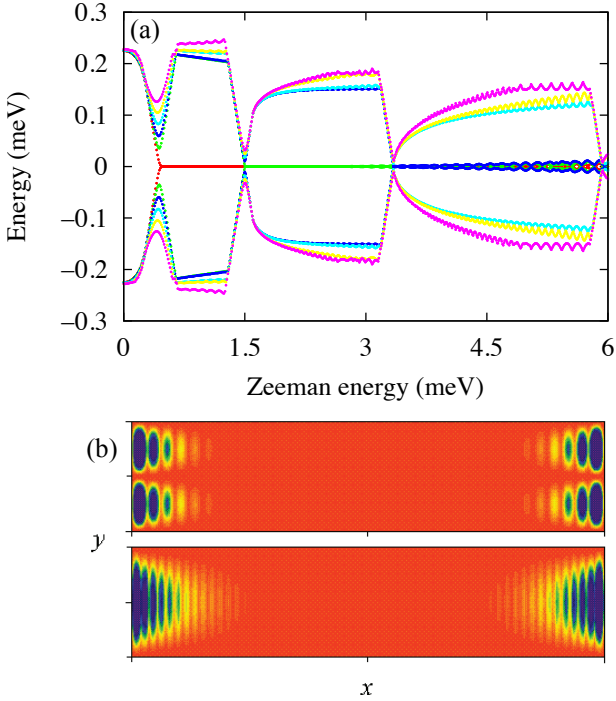,angle=-90,width=0.45\textwidth,clip}
}
\caption{(Color online) 
Results neglecting Rashba mixing, $\alpha_y=0$ in Eq.\ (\ref{rashba}),
and without magnetic orbital effects, for field orientation along $x$.
(a) Spectrum of eigenvalues as a function of Zeeman energy 
$\Delta_B\equiv g\mu_B B/2$. Only the twelve eigenvalues closer to zero are displayed;
higher or lower eigenvalues than the ones displayed correspond to
bulk excitations and are not shown. Parameters: $L_y=150\;{\rm nm}$, $L_x=3\;\mu{\rm m}$, 
$\Delta=0.225\;{\rm meV}$, $\mu=0$, $\alpha_x=0.045\;{\rm eV nm}$.
(b) Probability density for the 
Majorana-like edge modes corresponding to the second and first 
transverse modes for a Zeeman energy of $2\;{\rm meV}$ in panel (a).}
\label{fig2}
\end{figure}

\emph{The model.} We consider a semiconductor nanowire with a 
strong Rashba SO and proximity-induced pairin in the presence of a magnetic field $B$ with a given orientation as sketched in Fig.\ \ref{fig1}. 
The model Hamiltonian can be written as 
 $\mathcal{H} _{NW}= \mathcal{H}_{SP} + \mathcal{H}_Z + \mathcal{H}_{SO}$,
where
\begin{eqnarray}
\mathcal{H}_{SP} &=& 
\frac{\Pi_x^2 + \Pi_y^2}{2m^{\ast}}  - \mu +V(x;L_x)+V(y;L_y)\;, \\ 
\mathcal{H}_Z &=& \frac{1}{2}g\mu_B\vec{B}\cdot\vec{\sigma}\;, \\
\mathcal{H}_{SO} &=& \frac{\alpha_R}{2} \left(\vec{\sigma} \times \vec{\Pi} \right)\cdot \vec{\cal E}\;,
\end{eqnarray}
with $m^*$ the effective mass, $\Pi_i = p_i + (e/c)A_i$ the canonical momentum ($e$ is the electron's 
charge, and $c$ is the speed of light) and $\vec{A}$ the vector potential.  $\mathcal{H}_{SP}$ is the single particle energy,
referred to the chemical potential $\mu$ with a hard-wall confinement in both $x$ 
and $y$ directions given by $V(a;L)=0$ for $0<a<L$ and $V(a;L)=\infty$ 
otherwise. $\mathcal{H}_Z$ corresponds to the Zeeman term with $\Delta_B\equiv g\mu_B B/2$.  
$\mathcal{H}_{SO}$
is the Rashba SO Hamiltonian. The constant vector $\vec{\cal E}$ in $\mathcal{H}_{SO}$ corresponds 
to the effective electric field due to confinement. In our planar geometry (Fig.\ \ref{fig1}), typical 
of Rashba systems, the 
strongest confinement  occurs along the $z$ axis and, therefore, $\mathcal{H}_{SO} \approx \alpha/\hbar\left(\sigma_x \Pi_y - \sigma_y \Pi_x \right)$ 
with $\alpha=\alpha_R {\cal E}_z /2$. 
In a convenient gauge ($\vec{A}=-B_z y \hat{u}_x$) $\Pi_x=p_x-e yB_z/c$ and $\Pi_y=p_y$. 
Notice then that both $\mathcal{H}_{SP}$ and $\mathcal{H}_{SO}$ are modified by the magnetic orbital effects, represented by the magnetic length $\ell_z=\sqrt{\hbar c/eB_z}$. 
More specifically, in the case of the Rashba Hamiltonian we have
\begin{equation}\label{rashba}
\mathcal{H}_{SO}=\frac{\alpha_x}{\hbar} p_x \sigma_y-\frac{\alpha_y}{\hbar} p_y \sigma_x-\frac{\alpha_x}{\hbar}\frac{y\sigma_y}{\ell_z^2}\,,
\end{equation}
where the distinction between $\alpha_x$ and $\alpha_y$ is introduced for later convenience. 
Clearly, in Rashba NWs $\alpha_x=\alpha_y=\alpha$, however
situations with $\alpha_y\neq \alpha_x$,
including the case $\alpha_y=0$ have been theoretically proposed 
in NWs without spin orbit interaction using inhomogeneous fields $B(x)$
or spatially modulated g-factors. \cite{Flensberg11}
In general, Eq. (\ref{rashba}) mixes different quantum well subbands in $x$ (through the $p_x$ operator) 
and in $y$ ($p_y$ and $y$ operators).
Since $L_x\gg L_y$ the relevant mixing contribution is 
$\alpha_y p_y \sigma_x$ which we termed as Rashba mixing.  
Finally, we stress that orbital effects modify
the SO coupling, including the Rashba mixing. 
As shown below, they alter dramatically the topological phases of multiband NWs.

The second-quantized Hamiltonian can be written as $\mathcal{H}_{NW} \to \sum_{n,n'}  
\sum_{\sigma,\sigma'} \langle n\sigma|\mathcal{H}_{NW}|n' \sigma'\rangle c_{n\sigma}^{\dagger}c_{n'\sigma'}$ 
where $n\equiv\{n_x,n_y\}$ are the quantum numbers associated with the transverse modes 
due to confinement and $\sigma=\uparrow,\downarrow$ denotes the spin. 
When the NW is proximity-coupled to an ordinary s-wave superconductor the
BCS-Hamiltonian is taken into account, $\mathcal{H}_{SC} = \sum_{n}\left[ \Delta\, c_{n\uparrow}^{\dagger}c_{n \downarrow}^{\dagger} + {\mathit h.c.}\right]$. 
One then obtains the low-energy physics of this system in a Bogoliubov-deGennes description. In matrix form,
$\mathcal{H}_{BdG} = 
\frac{1}{2}\sum_{n,n'}\Psi_{n}^{\dagger}\mathcal{H}_{BdG}^{(nn')} \Psi_{n'}$ where
$\Psi_n=(c^\dagger_{n\uparrow},c^\dagger_{n\downarrow},c_{n\downarrow},-c_{n\uparrow})^\dagger$
and
\begin{eqnarray}
\label{eqbdg}
\mathcal{H}_{BdG}^{(nn')} = 
\begin{pmatrix}
\mathcal{H}^{nn'}_{NW}         & \Delta \\
\Delta^{\ast} & i\sigma_y[\mathcal{H}^{nn'}_{NW}]^{\ast} i\sigma_y
\end{pmatrix}
\;.
\end{eqnarray}
The results shown below are obtained by exact numerical diagonalization of Eq.\ (\ref{eqbdg}) 
and confirm the emergence of NW gapped low-energy eigenstates for some range of parameters. 

The existence of gapped zero-energy modes is the signature of Majorana physics. 
Let us label positive energy states $E_I$ by an index $I=1,2,\dots$ in increasing
energy order. Analogously, negative energy states in decreasing energy order
are labelled by $I=-1, -2, \dots $. In the Dirac picture of fermions, positive-energy states are particle states 
while negative ones are their conjugated or antiparticle ones. Clearly,
when $|E_I|$ is sizeable non-zero  $|I\rangle$ 
and $|-I\rangle$ are different stationary eigenstates of ${\cal H}_{BdG}$. 
We can now form the two combinations
\begin{eqnarray}
|\gamma_a^{(I)}\rangle &=& \frac{1}{\sqrt{2}}(|I\rangle+|-I\rangle),
\\
|\gamma_b^{(I)}\rangle &=& \frac{i}{\sqrt{2}}(|I\rangle-|-I\rangle)\;.
\end{eqnarray}
These are the Majorana states that, in general, are not
 eigenstates of ${\cal H}_{BdG}$ (see Eq. 4) unless
$E_I\approx E_{-I}\approx 0$.
These states correspond to zero modes 
separated by a sizeable energy gap from the rest of the spectrum. 
In the following we will carefully investigate the robustness of this behavior
in finite samples and the role of the Rashba mixing term.

\emph{Ocurrence of Majorana modes in finite nanowires.} It is estimated that in infinite strictly 1D quantum wires the occurrence of a Majorana pair requires
sufficiently large Zeeman energies at magnetic fields that exceed the critical value 
$B_c$ given by
$g\mu_B B_c/2=\sqrt{\mu^2+\Delta^2}$. \cite{PhysRevLett.105.077001,PhysRevLett.105.177002} 
In the quasi 1D  case one has a similar condition for the critical field $B_c(n_{y})$ at which
a  Majorana pair emerges from the $n_y$ transverse mode: $g\mu_B B_c(n_y)/2=\sqrt{(\varepsilon_{n_y}-\mu)^2+\Delta^2}$ 
 where $\varepsilon_{n_y}$ is the transverse mode energy. This scenario is confirmed in   Fig.\ \ref{fig2}(a) where
we show the spectrum when the Rashba mixing term is absent and the magnetic field is applied along $\hat{x}$ direction. 
Remarkably, by increasing the magnetic field successive gapped low-energy pairs, each belonging to a different transverse mode, appear in the spectrum.
We emphasize that Fig.\ \ref{fig2}(a) corresponds to a situation in which transverse modes are \emph{uncoupled} because $\alpha_y=0$.  To better illustrate the edge character of the Majorana pairs
we have  plotted  in Fig.\ \ref{fig2}(b) the probability density for the the first and second 
transverse modes for $\Delta_B=2\;{\rm meV}$. 
Remarkably, the pair energies of the spectrum are not exactly zero, but oscillate
around zero with increasing $B$, showing a steadily increasing amplitude. 
This is a finite size effect, indicating that the shortness of the wire eventually dominates
removing Majorana pairs from zero energy and destroying the gap with the 
nearby states. 

\begin{figure}[h]
\centerline{
\epsfig{file=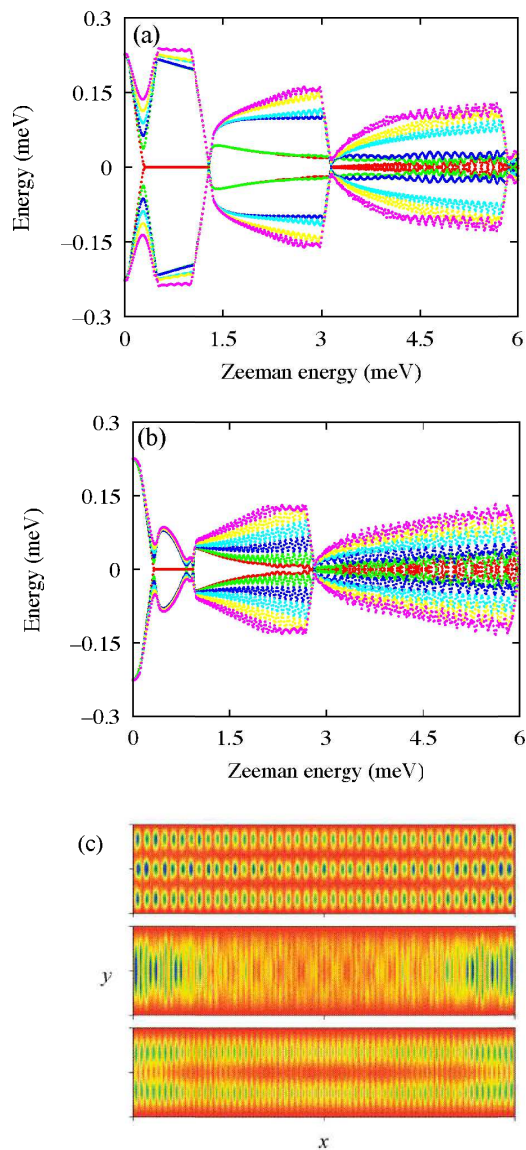,angle=-90,width=0.4\textwidth,clip}
}
\caption{(Color online) 
Same as Fig.\ \ref{fig2} including Rashba mixing. The same
parameters have been used, except for $\alpha_y=0.6\alpha_x$ in panel (a)
and $\alpha_y=\alpha_x$ in panel (b).
The densities of panel (c) are for the three lowest modes when $\Delta_B=4\;{\rm meV}$.}
\label{fig3}
\end{figure}

The value of the Rashba coupling assumed in Fig.\ 2 
is within the range of typical values for InAs-based nanostructures.
In the present context,  however, this coupling could be affected by the
nearby superconductor. In first approximation we have neglected this 
variation. Anyway, we have also checked that the physics we describe 
remains unaffected when $\alpha$ is modified by a factor 2.

The effect of the Rashba mixing on the occurrence of multiple Majorana 
pairs is shown in Fig.\ \ref{fig3}. When Rashba mixing is present we find \emph{a single pair of Majorana modes}  
in the low magnetic field regime. The Rashba mixing destroys the coexistence among
different Majorana pairs at moderate magnetic fields. This is clearly illustrated in 
Fig.\ \ref{fig3} where the spectrum is shown for 
$\alpha_y=0.6\alpha_x$ [Fig.\ \ref{fig3}(a)], and $\alpha_y=\alpha_x$ [Fig.\ \ref{fig3}(b)]. 
In both cases, the Rashba mixing leads to an effective coupling between 
low energy modes and dramatically affects
the spectrum in the regions where two or three Majorana pairs would coexist.
Because of the Rashba mixing, a clear gap region around zero energy appears [see Fig.\ \ref{fig3}(b) for $1$ meV $\gtrsim\Delta_B\gtrsim 3$ meV].
By increasing further the magnetic field, there are no visible Majorana pairs of the spectrum
 in both Figs.\ \ref{fig3}(a), and (b).  In these cases the eigenvalue spectrum becomes dense,
trivial in topological language, due to the Rashba mixing.
Therefore, the effect of Rashba mixing on the occurrence 
of multiple Majorana modes is two-fold:
 their zero energy character is lifted and their gap protection from the 
rest of the states tends to vanish. 
 Finally in Fig.\ \ref{fig3}(c) we show the spatial probability distributions 
of the three lowest eigenstates at large
magnetic fields ($\Delta_B=4$ meV). In this case the Rashba mixing clearly
destroys the edge-mode character of the states. 

For completeness, we have investigated the robustness 
of the single Majorana regime when the chemical potential is changed (not shown here).
We find an alternating trivial and nontrivial behavior 
as $\mu$ increases \cite{PhysRevB.83.184520} due to a sequence 
of occupied and unoccupied  single transverse modes. 
Notice that single Majorana pairs do not necessarily emerge from
the $n_y=1$ transverse mode but actually from the lowest energy occupied transverse mode.  
A good strategy towards detecting Majorana modes is to use magnetic fields 
 close to the critical field when only a single transversal mode contributes to the formation  of a
 Majorana pair. In such situation, the sequence  of occupied 
and unoccupied modes, as one increases the
chemical potential, will produce a sequence of regions with
a single Majorana pair each, so that Rashba mixing is
not efficient. We emphasize  that we do not find
such alternating behavior between trivial/nontrivial
phases when we vary the Zeeman field because of the Rashba mixing
and the size effects. Nevertheless, the sequence of trivial/nontrivial phases in the whole ($\mu$, $\Delta_B$) 
phase space \cite{PhysRevB.83.184520,PhysRevB.84.144522,PhysRevLett.106.127001,Fazio} 
should be recovered for  $L_x/L_y\to\infty$, when topological arguments become exact. From this point 
of view, our results demonstrate that the extrapolation of the topological
phase diagram to finite samples with $L_x/L_y\approx20$ is not sufficiently justified.  

\begin{figure}[t]
\centerline{
\epsfig{file=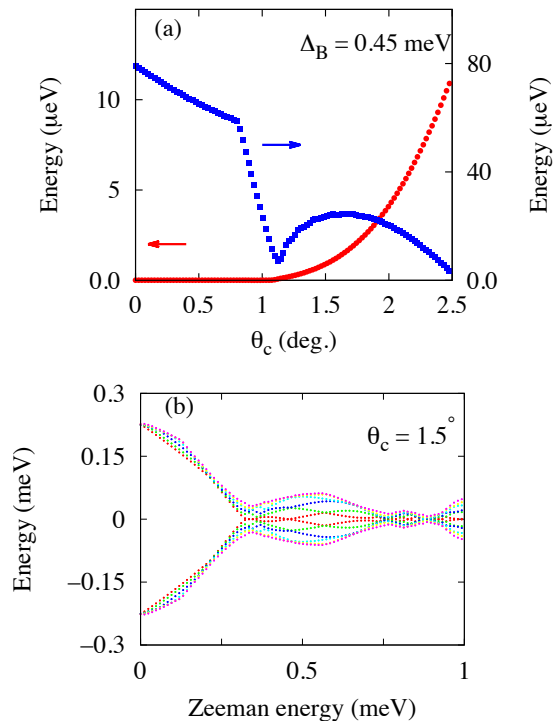,angle=-90,width=0.4\textwidth,clip}
}
\caption{(Color online) 
a) Energy of the lowest eigenvalue as a function of the field tilting
angle (circles) and of the gap with the next state (squares).
We have used a Zeeman energy of 0.45 meV.
b) Same as Fig.\ \ref{fig2} for a field 
tilting angle of 1 degree out of the plane.
Same parameters of Fig.\ \ref{fig3}b.
}
\label{fig4}
\end{figure}

\emph{Magnetic orbital effects.} All the above results were obtained assuming a magnetic field with 
a perfect in-plane orientation. We finish this discussion by studying the 
role of magnetic orbital effects, due to $B_z\neq0$. The polar angle $\theta_c=\tan^{-1} B_z/B_x$
quantifies a small out-of-plane deviation of the magnetic field.
Our results are shown in Fig.\ \ref{fig4}. 
A  small vertical component of the field 
is sufficient to steadily increase the energy  
of the lowest mode, from a clear  
Majorana-like character with $E_1\approx0$ for $\theta_c=0^{\rm o}$   
to $E_1\approx 0.3\Delta$ for $\theta_c=2.5^{\rm o}$ [see Fig. 4(a)]. 
At the same time, the gap from the lowest to the next
state, $E_2-E_1$, is reduced as the magnetic field deviation from the wire plane
decreases. Thus, magnetic orbital effects suppress the Majorana character of the low-energy modes.  
This is demonstrated in Fig.\ \ref{fig4}(b), where just few degrees of tilting are 
enough to destroy the low energy modes
as a function of $B$.

\emph{Conclusions.} Summarizing, we have investigated the effect of the Rashba intermode coupling in \emph{finite
multiband} semiconductor nanowires when superconductivity is proximity induced
and in the presence of a magnetic field. We find that Majorana physics appears provided
that only one single transverse mode leads to the Majorana pair formation.  
The coexistence of multiple Majorana pairs is excluded by the presence of the Rashba mixing that
hybridizes Majorana pairs and reduces their gap protection from the rest of the states. 
 Additionally, we have studied how magnetic orbital effects affect the formation of gapped
 zero energy modes.  We conclude that even in the single Majorana mode regime, magnetic
 orbital effects lead to the destruction of those modes. 
 While the fragility against magnetic orbital effects seems ultimately 
 unavoidable in a planar geometry, the hybridization of multiple Majorana pairs 
 could be eliminated as it has been shown in a recent proposal in Ref.\ [\onlinecite{Flensberg11}], where Majorana physics 
occurs in nanowires without Rashba coupling. Here, Majorana modes are originated by either the NW curvature or 
by an inhomogeneous Zeeman field generating effective band-diagonal coupling.

\emph{Note added---} During the last stages of this work the experimental evidence of Majorana modes in 
hybrid InSb nanowire has been presented. \cite{mou12,den12} Although our geometry (planar) differs from
the  experimental one (cylindrical), we have computed Fig.\ 2 and Fig.\ 3 for the experimental values corresponding to
 InSb nanowires  ($\Delta=0.23$ meV, $\alpha=0.023$ eVnm) showing
that the magnetic instability reported here (the splitting at large magnetic fields in Fig. 3b)
agrees with the experiment.\cite{mou12}  Nevertheless, the fragility predicted here against orbital effects has
not been observed in the experiment when tilting the magnetic field in
the vertical direction with respect to the cylinder axis. We believe this is
due to the different role of tilting in planar and
cylindrical geometries. A precise investigation of this mechanism is
in progress.

\emph{Acknowledgements.} We thank D. S\'{a}nchez for a critical 
reading of the manuscript and illuminating discussions. This work 
was supported by Grants No. FIS2008-00781, FIS2009-08744, 
FIS2011-23526, and CSD2007-00042 (CPAN) 
of the Spanish Government.


\end{document}